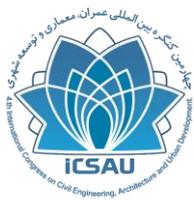

# Output-only Modal Identification of beams with different boundary condition

M.R. Davoodi[1], S.A. Mostafavian[2], S.R. Nabavian[3], GH.R. Jahangiry[4*]

1. Associate Professor of Civil Engineering, Babol University of Technology, Iran (davoodi@nit.ac.ir)
2. Assistant Professor of Civil Engineering, Payame Nor University, Sari branch, Iran (a.mostafavian@stu.nit.ac.ir)
3. Ph.D Student of Civil Engineering, Faculty Member, Tabari University of Babol, Iran (nabavian@stu.nit.ac.ir)
4. M.sc. Student of Civil Engineering, Tabari University of Babol, Iran (gr.jahangiry5011@gmail.com)

## Abstract

Structural Health Monitoring (SHM) evaluates the integrity of a structure by observing its dynamic responses by an array of sensors over time to determine the current health state of the structure. The most important step of SHM is system identification that in civil structures is identification of modal parameters of structures. Due to numerous limitations of input-output methods, system identification of ambient vibration structures using output-only identification techniques has become a key issue in structural health monitoring and assessment of engineering structures. In this paper, four beams with different boundary condition and with arbitrary loading have been modeled in finite element software, ANSYS, and the responses (Acceleration of nodes) have been achieved. By using these data and the codes written in MATLAB software, the modal parameters (natural frequencies, mode shapes) of the beams are identified with FDD (frequency Domain Decomposition) and PP (Peak Picking) methods and then justified with the results of input-output method which was determined by frequency relation function (FRF). The results indicate that there is a good agreement between three methods for determining dynamic characteristics of beams.

## 1. Introduction

Identification of the global dynamic properties of civil engineering structures using vibration responses is a necessary step in several types of analysis including, for instance, model updating and structural health monitoring (SHM). Ageing and structural deterioration are also crucial issues in structural design and maintenance. The regular identification of modal parameters can play a relevant role in the development of effective structural health monitoring systems. Over the last thirty years, these circumstances have led civil engineers to exploit a number of techniques developed in the fields of system identification and experimental modal analysis [1]. These techniques first referred to electrical engineering applications, but they progressively spread to other fields including automotive, aerospace and civil engineering. With input-output modal identification procedures, such techniques enable the experimental identification of the dynamic properties of structures. Traditional experimental modal analysis (EMA), however, suffers from several limitations, as described below.

– It requires artificial excitation to evaluate frequency response functions (FRF) or impulse response functions (IRF). In some cases, such as civil structures, providing adequate excitation is difficult if not impossible [2].

– Operational conditions are often different from those adopted in tests because traditional EMA is conducted in a laboratory environment.

– The boundary conditions are simulated because tests are usually conducted in a laboratory environment on components instead of with complete systems.

As a consequence, since the early 1990s, the civil engineering community has paid an increasing amount of attention to OMA, with applications for several structures, including buildings, bridges and offshore platforms. OMA uses structural response measurements from ambient excitation to extract modal characteristics. Thus, it is also called ambient, natural-excitation or output-only modal analysis. Compared to traditional EMA, OMA is attractive due to a number of advantages:

– It is faster and cheaper than EMA;
– No excitation equipment or boundary condition simulations are needed;
– It does not interfere with the normal use of the structure;
– It enables the identification of modal parameters that are representative of the entire system under actual service conditions; and
– OMA also can be used for vibration-based structural health monitoring and to detect damage in structures [3].

Modal parameters of the ambient vibration structures consist of natural frequencies, mode shapes and modal damping ratios. So far, a number of mathematical models on the output-only identification techniques have been developed and roughly classified by either parametric methods in the time domain or nonparametric ones in the frequency domain [4]. Each identification method in either the time domain or the frequency one has its own advantage and limitation. If a model is fitted to data, the technique is parametric. These techniques are more complex and computationally demanding, and they usually perform better than the faster and easier non-parametric techniques, which, however, are preferred for initial insight into the identification problem. Generally, parametric methods such as Ibrahim time domain (ITD), Eigensystem realization algorithm (ERA) or Random decrement technique (RDT) are preferable for estimating modal damping but difficulty in natural frequencies, mode shapes extraction, whereas nonparametric ones such as Peak-picking (PP), Frequency domain decomposition (FDD) or Enhanced frequency domain decomposition (EFDD) advantage on natural frequencies, mode shapes extraction, but uncertainty in damping estimation [5], [6]. Recently, new approach based on Wavelet transform (WT) and Hilbert-Huang transform (HHT) have been developed for output-only identification techniques in the time-frequency plane. The most undemanding method for output-only modal parameter identification is the basic frequency domain (BFD) technique [7], also called the Peak-Picking method, because the identification of eigenfrequencies is based on peak picking in the power spectrum plots. However, this method can lead to erroneous results if the basic assumptions of low damping and well-separated frequencies are not fulfilled. In fact, the method identifies the operational deflection shapes, which are the superpositions of multiple modes for closely spaced modes. The singular value decomposition (SVD) of the power spectral density (PSD) matrix overcomes these shortcomings and leads to the frequency domain decomposition (FDD) method [8], which is capable of detecting mode-multiplicity. However, both of these techniques are non-parametric methods because the modal parameters are obtained without fitting a mathematical model to the measured data.

In this article, beams with different boundary conditions under ambient vibration are studied and the natural frequencies and the mode shapes of beams are determined using PP and FDD methods.

## 2. Peak-Picking

The peak-picking method (PP) is the simpler and more practical method for modal identification. In spite of some drawbacks, the PP method gives very fast results and is useful as a pre-process tool when dynamic monitoring is performed. The PP method was systematized by Felber [9]. In this method the natural frequencies of the structures are determined as the peaks of the Average Normalized Power-Spectral Densities (ANPSDs) [10], the damping factors are determined using the Half Power Bandwidth Method [11] , and the components of the mode shapes are determined by the values of the transfer functions at the natural frequencies [12]. The main limitations of the PP method, is that picking the peaks is often a subjective task, identifying close frequencies is difficult, spurious modes can be confused as real ones, operational deflection shapes are obtained instead of mode shapes and the damping estimates are unreliable [12].

Felber introduced a new function instead of PSD function in order to determine the frequencies. This function called the Averaged Normalized Power Spectral Density (ANPSD) is defined as the average of a group of $l$ normalized power spectral densities (NPSDs).

The ANPSD functions are calculated using:

$$\text{ANPSD}(f_k) = \frac{1}{l} \sum_{i=1}^{i=l} NPSD_i(f_k)$$

Where $NPSD_i(f_k)$ is defined as

$$NPSD_i(f_k) = \frac{PSD_i(f_k)}{\sum_{k=0}^{k=n} PSD_i(f_k)}$$

And $f_k$ is the kth discrete frequency and n is the number of discrete frequencies.

## 3. Frequency Domain Decomposition

Among the nonparametric methods in the frequency domain, FDD has been very widely used recently for output-only system identification through the ambient vibration measurements due to its reliability, straightforward and effectiveness [13], applied for wind-excited structures [6]. FDD is also powerful for closed natural frequencies extraction. However, FDD always requires the prior-selected natural frequencies as well as its strict hypotheses of uncorrelated white noise excitations and lightly structural damping. Under these strict hypotheses, the output PSD matrix can be expressed similarly as form of conventionally matrix decompositions, consequently, first-order linear approximation of the output PSD matrix is used for estimating mode shapes and damping. Output PSD matrix can be decomposed via fast-decayed decomposition methods such as QR decomposition and almost Singular value decomposition (SVD). By doing so, the spectral densities functions are decomposed in the contributions of different modes of a system that, at each frequency, contribute to its response. From the analysis of the singular values it is possible to identify the auto power spectral density functions corresponding to each mode of a system. In the FDD method, the mode shapes are estimated as the singular vectors at the peak of each auto power spectral density function corresponding to each mode. The FDD technique is an extension of the BFD method. The theoretical basis can be summarized as follows. The relationship between the input *x(t)* and the output *y(t)* can be written in the following form [8]:

$$[G_{yy}(\omega)] = [H(\omega)]^*[G_{xx}(\omega)][H(\omega)]^T$$

Where $G_{xx}(\omega)$ is the $r \times r$ input PSD matrix; $r$ is the number of inputs; $G_{yy}(\omega)$ is the $m \times m$ output PSD matrix; $m$ is the number of outputs; $[H(\omega)]$ is the $m \times r$ FRF matrix; and the superscripts $^*$ and $^T$ denote complex conjugate and transpose, respectively.

The FRF matrix can be expressed in a typical partial fraction form, which is used in classical modal analysis, in terms of poles, *l*, and residues, *[R]*:

$$[H(\omega)] = \frac{[Y(\omega)]}{[X(\omega)]} = \sum_{k=1}^{n}\left(\frac{[R_k]}{j\omega - \lambda_k} + \frac{[R_k]^*}{j\omega - \lambda_k^*}\right)$$

with

$$\lambda_k = -\sigma_k + j\omega_{dk}$$

Where *n* is the number of modes; $\lambda_k$ is the pole of the *k*th mode; $\sigma_k$ is the modal damping decay constant; and $\omega_{dk}$ is the damped natural frequency of the *k*th mode. $[R_k]$ is the residue, and it is given by

$$[R_k] = \{\phi_k\}\{\gamma_k\}^T$$

Where $\{\phi_k\}$ is the mode shape vector, and $\{\gamma_k\}$ is the modal participation vector. Therefore, combining eq. (1) and (2) and assuming that the input is random in both time and space and has a zero mean white noise distribution (i.e., the PSD is constant: $[G_{xx}(\omega)] = [C]$), the output PSD matrix can be written as

$$[G_{yy}(\omega)] = \sum_{k=1}^{n}\sum_{s=1}^{n}\left[\frac{[R_k]}{j\omega - \lambda_k} + \frac{[R_k]^*}{j\omega - \lambda_k^*}\right][C]\left[\frac{[R_k]}{j\omega - \lambda_k} + \frac{[R_k]^*}{j\omega - \lambda_k^*}\right]^H$$

Using the Heaviside partial fraction theorem for polynomial expansions, the following result can be obtained:

$$[G_{yy}(\omega)] = \sum_{k=1}^{n}\left(\frac{[A_k]}{j\omega - \lambda_k} + \frac{[A_k]^*}{j\omega - \lambda_k^*} + \frac{[B_k]}{-j\omega - \lambda_k} + \frac{[B_k]^*}{-j\omega - \lambda_k^*}\right)$$

This is the pole-residue form of the output PSD matrix. $[A_k]$ is the $k^{th}$ residue matrix of the output PSD; it is an $m \times m$ hermitian matrix given by

$$[A_k] = [R_k][C]\sum_{s=1}^{n}\left(\frac{[R_s]^H}{-\lambda_k - \lambda_s^*} + \frac{[R_s]^T}{-\lambda_k - \lambda_s}\right)$$

If only the $k^{th}$ mode is considered, the following contribution is obtained:

$$[A_k] = \frac{[R_k][C][R_k]^H}{2\sigma_k}$$

This term can become dominant if the damping is low, and a residue proportional to the mode shape vector can be obtained as follows:

$$[A_k] \propto [R_k][C][R_k]^H = \{\phi_k\}\{\gamma_k\}^T[C]\{\gamma_k\}\{\phi_k\}^T = d_k\{\phi_k\}\{\phi_k\}^T$$

Where $d_k$ is a scaling factor for the $k^{th}$ mode.
For a lightly damped system in which the contribution of the modes at a particular frequency is limited to a finite number (usually one or two), the response spectral density matrix can be written in the following final form:

$$[G_{yy}(\omega)] = \sum_{k \in Sub(\omega)}\left(\frac{d_k\{\phi_k\}\{\phi_k\}^T}{j\omega - \lambda_k} + \frac{d_k^*\{\phi_k\}^*\{\phi_k\}^{*T}}{j\omega - \lambda_k^*}\right)$$

Where $k \in Sub(\omega)$ is the set of contributing modes at the considered frequency. The SVD of the output PSD matrix known at discrete frequencies $\omega = \omega_i$ gives

$$[\hat{G}_{yy}(j\omega_i)] = [U]_i[S]_i[U]_i^H$$

Where the matrix $[U]_i$ is a unitary matrix holding the singular vector $\{u_{ij}\}$, and $[S]_i$ is a diagonal matrix holding the scalar singular values $s_{ij}$. Near a peak corresponding to the $k^{th}$ mode in the spectrum, this mode will be dominant. If only the $k^{th}$ mode is dominant, only one term in eq. (10) exists, and the PSD matrix approximates to a rank one matrix:

$$\hat{G}_{yy}(j\omega_i) = s_i\{u_{i1}\}\{u_{i1}\}^H \omega_i \to \omega_k$$

In such a case, therefore, the first singular vector $\{u_{i1}\}$ represents an estimate of the mode shape:

$$\{\hat{\phi}\} = \{u_{i1}\}$$

And the corresponding singular value belongs to the auto power spectral density function of the SDOF system corresponding to the mode of interest. In the case of repeated modes, the PSD matrix rank is equal to the multiplicity number of the modes. The auto power spectral density function of the corresponding SDOF system is identified around the peak of the singular value plot by comparing the mode shape estimate $\{\hat{\phi}\}$ with the singular vectors associated with the frequency lines around the peak [14].

## 4. Numerical Modeling

First, four beams with different boundary conditions was modeled in finite element software, ANSYS, And the first five mode shapes has been achieved using modal analysis (Fig. 1).

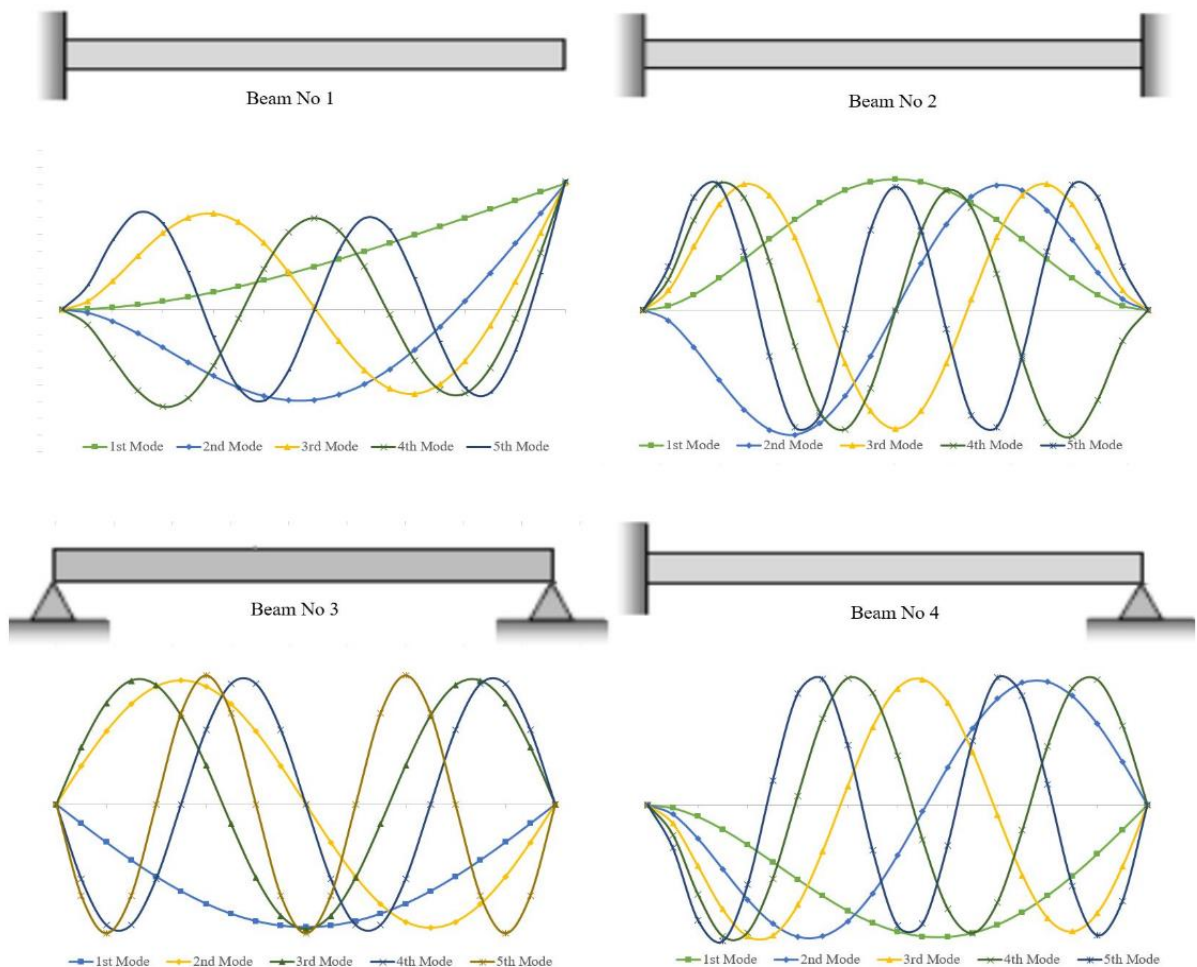

Fig. 1- First five mode shapes of the beams

Then the beams were subjected to arbitrary ambient vibration and the acceleration response of beams has been obtained. The PSD diagram for the acceleration response of cantilever beam (beam No.1) is drawn with PP and FDD methods (Fig. 2, Fig. 3).

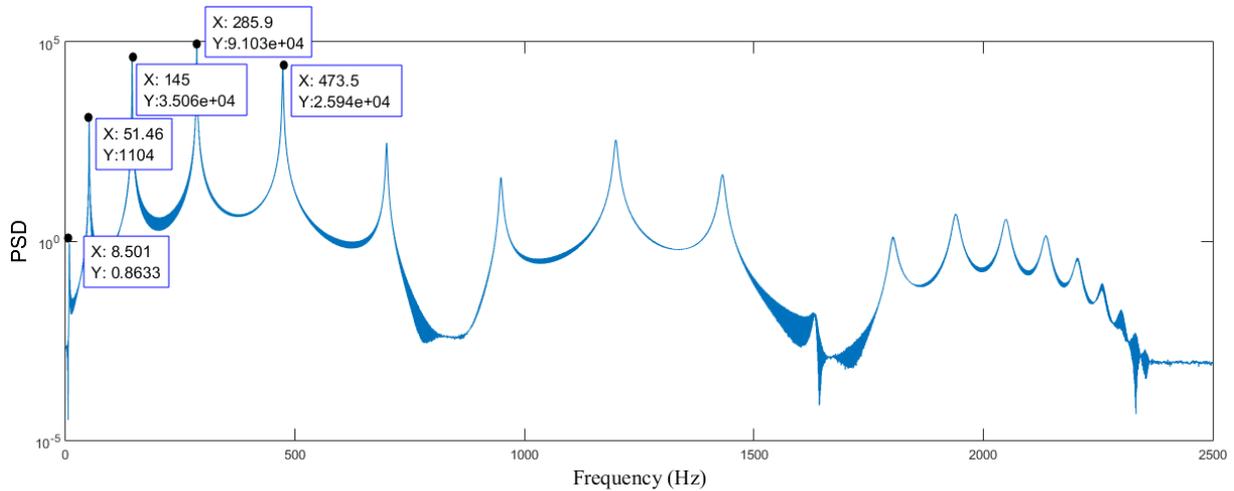

**Fig. 2- PSD plot of contilever beam (PP)**

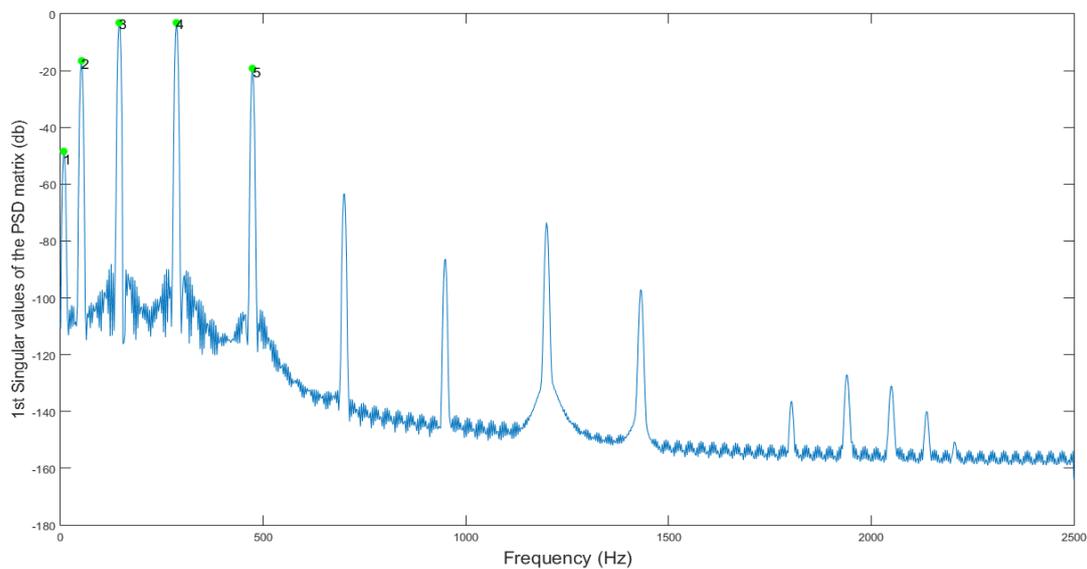

**Fig. 3- PSD plot of contilever beam (FDD)**

In the tables 1-4, the mode shape values (vertical deformation of the beams) for different beams with 10 elements has been shown with FEM, PP and FDD methods as well as their mode shapes (Fig. 4-7). According to the figures, there is a good agreement between the mode shapes drawn with different methods.

Table 1. Mode shape values for Beam No.1

| Method \ Node No. | 1 | 2 | 3 | 4 | 5 | 6 | 7 | 8 | 9 | 10 | 11 |
|---|---|---|---|---|---|---|---|---|---|---|---|
| Beam No.1 - Mode 1 | | | | | | | | | | | |
| FEM | 0 | 0.0164 | 0.0633 | 0.1357 | 0.2289 | 0.3385 | 0.4601 | 0.5899 | 0.7246 | 0.8619 | 1 |
| PP | 0 | 0.0169 | 0.0656 | 0.1411 | 0.2372 | 0.3476 | 0.4677 | 0.5947 | 0.7269 | 0.8627 | 1 |
| FDD | 0 | 0.0149 | 0.0599 | 0.1328 | 0.2288 | 0.3414 | 0.4644 | 0.5936 | 0.727 | 0.8627 | 1 |
| Mode 2 | | | | | | | | | | | |
| FEM | 0 | -0.0883 | -0.2963 | -0.5231 | -0.6835 | -0.7173 | -0.5966 | -0.3266 | 0.0599 | 0.5162 | 1 |
| PP | 0 | -0.0891 | -0.2987 | -0.527 | -0.6882 | -0.7221 | -0.6009 | -0.3302 | 0.0575 | 0.5148 | 1 |
| FDD | 0 | -0.0884 | -0.2962 | -0.5228 | -0.6828 | -0.7166 | -0.5962 | -0.327 | 0.0593 | 0.5158 | 1 |
| Mode 3 | | | | | | | | | | | |
| FEM | 0 | 0.2142 | 0.5983 | 0.7618 | 0.5389 | 0.0305 | -0.4755 | -0.6775 | -0.4297 | 0.1947 | 1 |
| PP | 0 | 0.2142 | 0.5981 | 0.7614 | 0.5382 | 0.03 | -0.4754 | -0.6767 | -0.4285 | 0.1957 | 1 |
| FDD | 0 | 0.214 | 0.598 | 0.7616 | 0.5389 | 0.0306 | -0.4752 | -0.6776 | -0.4299 | 0.1946 | 1 |
| Mode 4 | | | | | | | | | | | |
| FEM | 0 | -0.3614 | -0.7648 | -0.4621 | 0.3064 | 0.7357 | 0.3708 | -0.386 | -0.6935 | -0.1335 | 1 |
| PP | 0 | -0.3608 | -0.7635 | -0.4614 | 0.3055 | 0.7335 | 0.369 | -0.3855 | -0.6907 | -0.1309 | 1 |
| FDD | 0 | -0.3612 | -0.7645 | -0.4618 | 0.3064 | 0.7355 | 0.3705 | -0.386 | -0.6933 | -0.1332 | 1 |
| Mode 5 | | | | | | | | | | | |
| FEM | 0 | 0.514 | 0.71 | -0.1946 | -0.7639 | -0.0615 | 0.7437 | 0.3145 | -0.6258 | -0.448 | 1 |
| PP | 0 | 0.5158 | 0.714 | -0.1913 | -0.7631 | -0.0622 | 0.7437 | 0.3146 | -0.6275 | -0.4504 | 1 |
| FDD | 0 | 0.5148 | 0.7111 | -0.1949 | -0.7651 | -0.062 | 0.7445 | 0.3153 | -0.626 | -0.4448 | 1 |

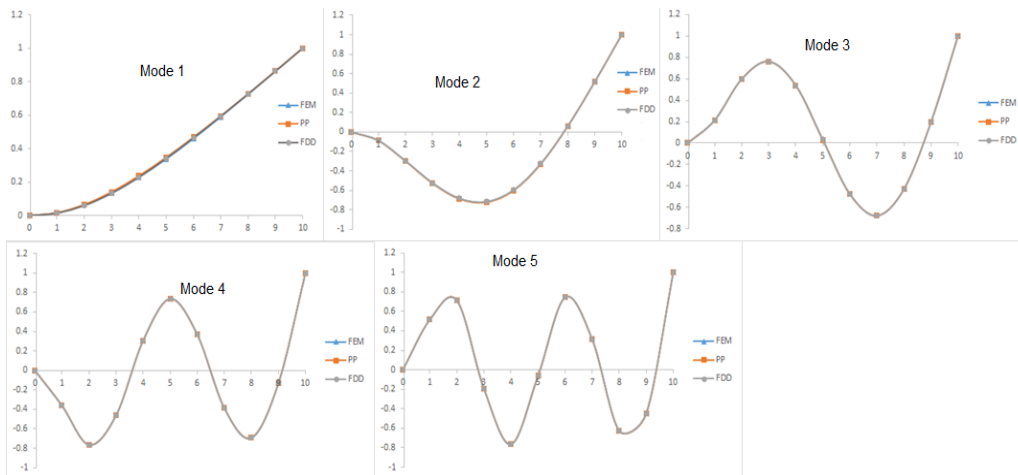

Fig. 4- Mode shapes of beam No.1

Table 2. Mode shape values for Beam No.2

| Method \ Node No. | 1 | 2 | 3 | 4 | 5 | 6 | 7 | 8 | 9 | 10 | 11 |
|---|---|---|---|---|---|---|---|---|---|---|---|
| Beam No.2 - Mode 1 | | | | | | | | | | | |
| FEM | 0 | 0.1142 | 0.3853 | 0.6874 | 0.9157 | 1 | 0.9157 | 0.6874 | 0.3853 | 0.1142 | 0 |
| PP | 0 | 0.1129 | 0.3819 | 0.683 | 0.9125 | 1 | 0.9189 | 0.6922 | 0.3891 | 0.1155 | 0 |
| FDD | 0 | 0.1138 | 0.3839 | 0.6854 | 0.9142 | 1 | 0.9169 | 0.6891 | 0.3866 | 0.1145 | 0 |
| Mode 2 | | | | | | | | | | | |
| FEM | 0 | 0.2846 | 0.7918 | 1 | 0.6901 | 0 | -0.6901 | -1 | -0.7918 | -0.2846 | 0 |
| PP | 0 | 0.2846 | 0.7917 | 1 | 0.69 | 0.0026 | -0.691 | -1.0012 | -0.7928 | -0.285 | 0 |
| FDD | 0 | 0.2846 | 0.7918 | 1 | 0.69 | 0.0002 | -0.6902 | -1.0002 | -0.792 | -0.2846 | 0 |
| Mode 3 | | | | | | | | | | | |
| FEM | 0 | 0.4772 | 1 | 0.5823 | -0.4308 | -0.9595 | -0.4308 | 0.5823 | 1 | 0.4772 | 0 |
| PP | 0 | 0.4772 | 1 | 0.5825 | -0.4298 | -0.9583 | -0.4303 | 0.5817 | 0.9991 | 0.4768 | 0 |
| FDD | 0 | 0.4772 | 1 | 0.5822 | -0.4305 | -0.9591 | -0.4305 | 0.5824 | 1 | 0.4772 | 0 |
| Mode 4 | | | | | | | | | | | |
| FEM | 0 | -0.6867 | -0.9214 | 0.3038 | 1 | 0 | -1 | -0.3038 | 0.9214 | 0.6867 | 0 |
| PP | 0 | -0.6823 | -0.9144 | 0.3066 | 1 | -0.00143 | -1.0007 | -0.3048 | 0.9207 | 0.6865 | 0 |
| FDD | 0 | -0.6852 | -0.9194 | 0.3042 | 1 | -0.000658 | -0.999562 | -0.3046 | 0.9203 | 0.6863 | 0 |
| Mode 5 | | | | | | | | | | | |
| FEM | 0 | 0.8174 | 0.4704 | -0.9473 | -0.1622 | 1 | -0.1622 | -0.9473 | 0.4704 | 0.8174 | 0 |
| PP | 0 | 0.8187 | 0.4718 | -0.9476 | -0.1627 | 1 | -0.1634 | -0.9485 | 0.471 | 0.8183 | 0 |
| FDD | 0 | 0.8172 | 0.4701 | -0.947 | -0.1619 | 1 | -0.1625 | -0.9473 | 0.4703 | 0.8172 | 0 |

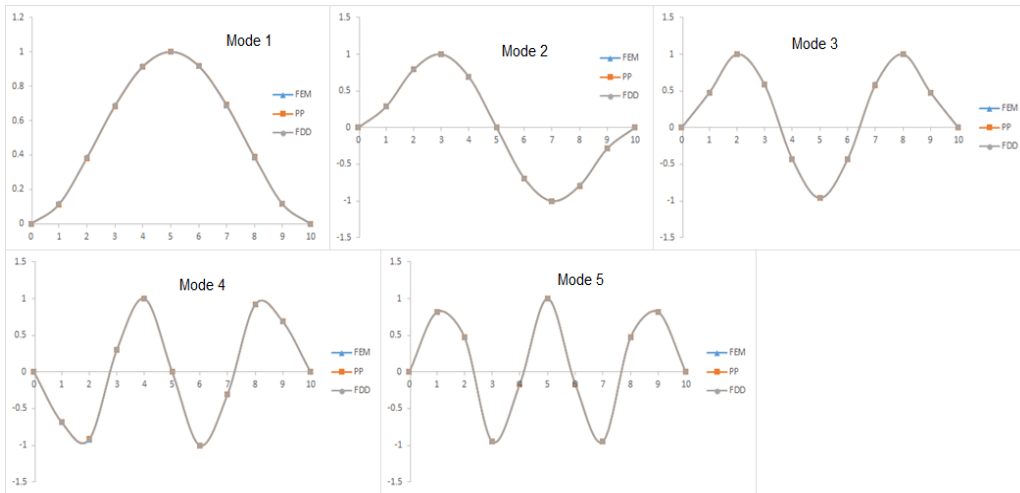

**Fig. 5- Mode shapes of beam No.2**

**Table 3. Mode shape values for Beam No.3**

| Node No. / Method | 1 | 2 | 3 | 4 | 5 | 6 | 7 | 8 | 9 | 10 | 11 |
|---|---|---|---|---|---|---|---|---|---|---|---|
| Beam No.3 - Mode 1 | | | | | | | | | | | |
| FEM | 0 | 0.309 | 0.5877 | 0.8089 | 0.951 | 1 | 0.951 | 0.8089 | 0.5877 | 0.309 | 0 |
| PP | 0 | 0.3059 | 0.5834 | 0.8054 | 0.9491 | 1 | 0.9529 | 0.8126 | 0.592 | 0.312 | 0 |
| FDD | 0 | 0.3062 | 0.5835 | 0.805 | 0.949 | 1 | 0.9514 | 0.809 | 0.5874 | 0.3088 | 0 |
| Mode 2 | | | | | | | | | | | |
| FEM | 0 | 0.618 | 1 | 1 | 0.618 | 0 | -0.618 | -1 | -1 | -0.618 | 0 |
| PP | 0 | 0.6182 | 1 | 0.9996 | 0.6179 | 0.0037 | -0.616 | -0.9975 | -0.998 | -0.6169 | 0 |
| FDD | 0 | 0.6181 | 1 | 1 | 0.6179 | 0.0002 | -0.6183 | -1.0002 | -1 | -0.6179 | 0 |
| Mode 3 | | | | | | | | | | | |
| FEM | 0 | -0.809 | -0.951 | -0.309 | 0.5878 | 1 | 0.5878 | -0.309 | -0.951 | -0.809 | 0 |
| PP | 0 | -0.8136 | -0.9571 | -0.3135 | 0.5857 | 1 | 0.5886 | -0.309 | -0.9524 | -0.8105 | 0 |
| FDD | 0 | -0.8094 | -0.9516 | -0.3093 | 0.588 | 1 | 0.5871 | -0.31 | -0.9514 | -0.8087 | 0 |
| Mode 4 | | | | | | | | | | | |
| FEM | 0 | -1 | -0.618 | 0.618 | 1 | 0 | -1 | -0.618 | 0.618 | 1 | 0 |
| PP | 0 | -0.9952 | -0.6141 | 0.6181 | 1 | 0.0024 | -0.9985 | -0.6177 | 0.6176 | 0.9996 | 0 |
| FDD | 0 | -1 | -0.6181 | 0.6179 | 1 | 0.0003 | -1 | -0.6181 | 0.6179 | 1 | 0 |
| Mode 5 | | | | | | | | | | | |
| FEM | 0 | -1 | 0 | 1 | 0 | -1 | 0 | 1 | 0 | -1 | 0 |
| PP | 0 | -0.9978 | 0.0019 | 1 | 0.0054 | -0.9977 | -0.0033 | 0.9974 | -0.0014 | -0.9984 | 0 |
| FDD | 0 | -1 | 0.0002 | 1 | 0.0006 | -1 | -0.0004 | 1 | -0.0002 | -1.0002 | 0 |

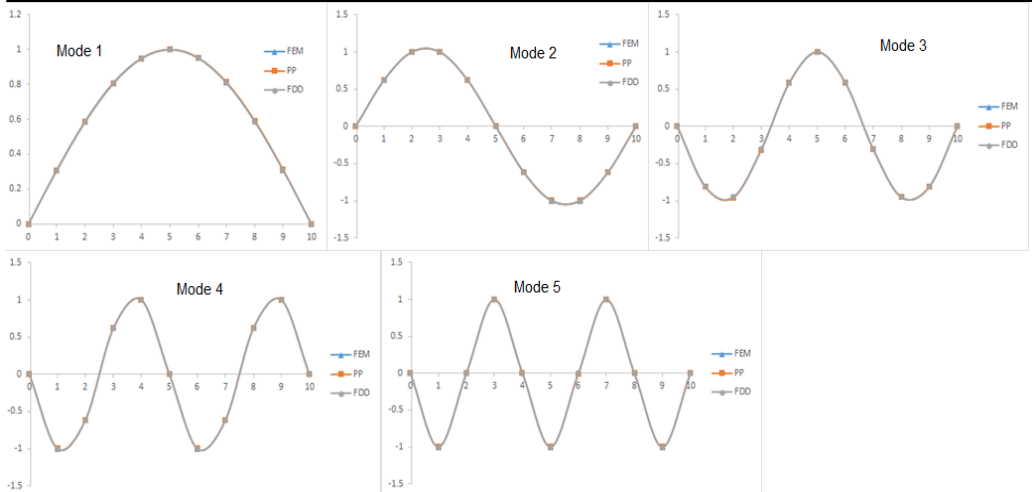

**Fig. 6- Mode shapes of beam No.3**

Table 4. Mode shape values for Beam No.4

| Node No. / Method | 1 | 2 | 3 | 4 | 5 | 6 | 7 | 8 | 9 | 10 | 11 |
|---|---|---|---|---|---|---|---|---|---|---|---|
| Beam No.4 - Mode 1 | | | | | | | | | | | |
| FEM | 0 | 0.0855 | 0.2985 | 0.5601 | 0.7992 | 0.9587 | 1 | 0.9072 | 0.6879 | 0.3705 | 0 |
| PP | 0 | 0.0847 | 0.2963 | 0.5578 | 0.7981 | 0.9587 | 1 | 0.906 | 0.6856 | 0.3687 | 0 |
| FDD | 0 | 0.0855 | 0.2987 | 0.5603 | 0.7996 | 0.9588 | 1 | 0.9066 | 0.687 | 0.37 | 0 |
| Mode 2 | | | | | | | | | | | |
| FEM | 0 | 0.2416 | 0.7103 | 1 | 0.8862 | 0.3832 | -0.2866 | -0.811 | -0.9426 | -0.6204 | 0 |
| PP | 0 | 0.2414 | 0.71 | 1 | 0.8866 | 0.3838 | -0.2867 | -0.8121 | -0.9444 | -0.6218 | 0 |
| FDD | 0 | 0.2414 | 0.7101 | 1 | 0.8863 | 0.3837 | -0.2862 | -0.811 | -0.9428 | -0.6207 | 0 |
| Mode 3 | | | | | | | | | | | |
| FEM | 0 | 0.4391 | 1 | 0.772 | -0.1425 | -0.9047 | -0.7966 | 0.0748 | 0.8752 | 0.8386 | 0 |
| PP | 0 | 0.4392 | 1 | 0.7714 | -0.1434 | -0.905 | -0.7958 | 0.0767 | 0.8771 | 0.8399 | 0 |
| FDD | 0 | 0.4392 | 1 | 0.772 | -0.1425 | -0.9048 | -0.7966 | 0.0748 | 0.8754 | 0.8387 | 0 |
| Mode 4 | | | | | | | | | | | |
| FEM | 0 | -0.6444 | -0.9907 | 0.0648 | 1 | 0.3954 | -0.8172 | -0.7746 | 0.4578 | 0.987 | 0 |
| PP | 0 | -0.6446 | -0.9911 | 0.0646 | 1 | 0.3952 | -0.8175 | -0.7745 | 0.4585 | 0.9878 | 0 |
| FDD | 0 | -0.6443 | -0.9903 | 0.0651 | 1 | 0.3952 | -0.8171 | -0.7743 | 0.4581 | 0.9874 | 0 |
| Mode 5 | | | | | | | | | | | |
| FEM | 0 | 0.7895 | 0.609 | -0.8455 | -0.4694 | 0.9218 | 0.3199 | -0.9737 | -0.1621 | 1 | 0 |
| PP | 0 | 0.7907 | 0.6141 | -0.8373 | -0.4648 | 0.9192 | 0.315 | -0.9752 | -0.163 | 1 | 0 |
| FDD | 0 | 0.7917 | 0.6111 | -0.8465 | -0.4694 | 0.9237 | 0.3188 | -0.9776 | -0.1634 | 1 | 0 |

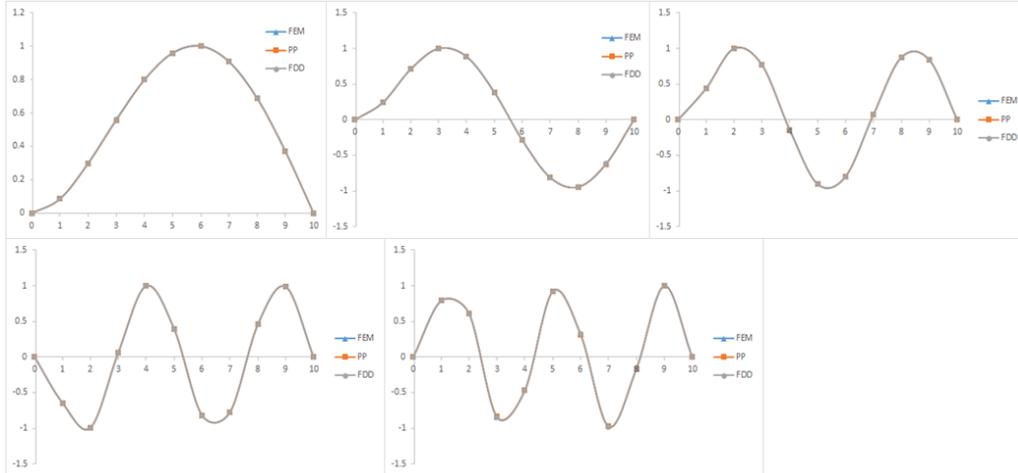

Fig. 7- Mode shapes of beam No.4

The following charts are related to MAC diagram for FEM and PP mode shapes and FEM and FDD mode shapes (Fig. 8-11).

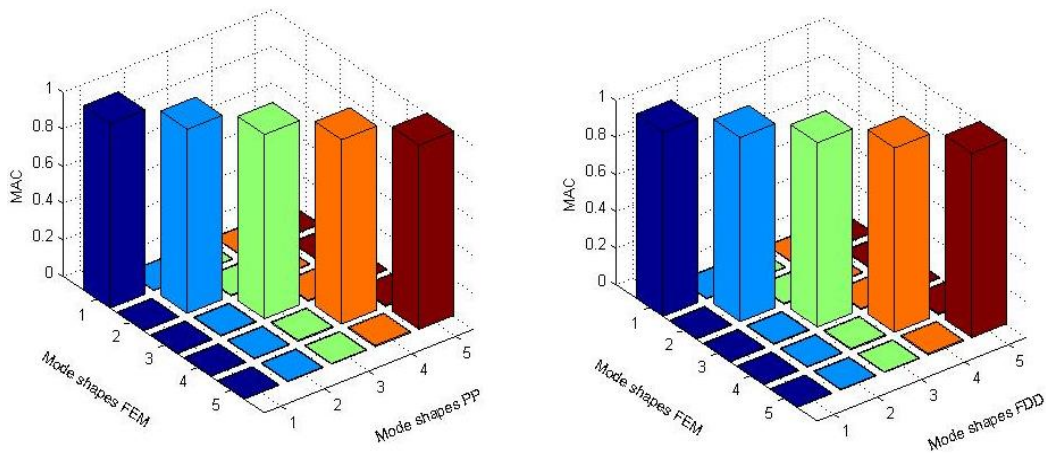

Fig. 8- MAC diagram of beam No.1

**4th. International Congress on Civil Engineering , Architecture
and Urban Development
27-29 December 2016, Shahid Beheshti University , Tehran , Iran**

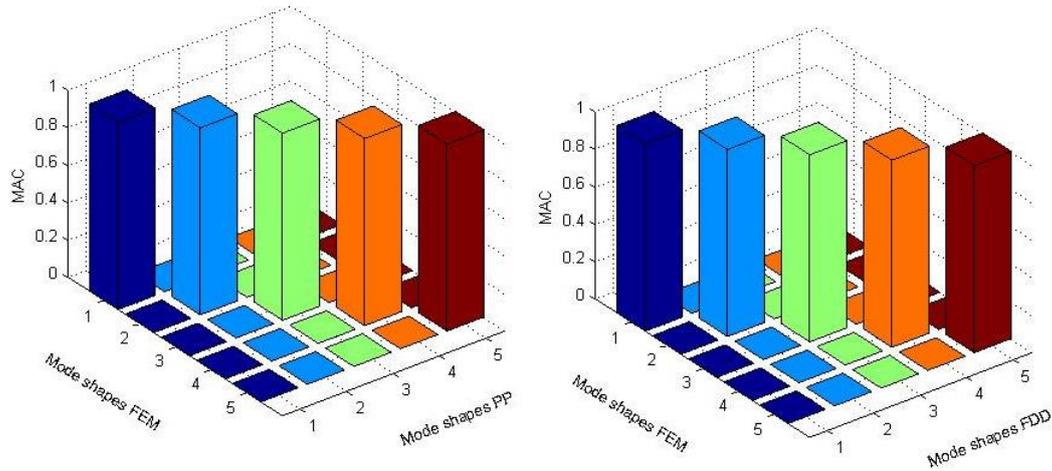

**Fig. 9- MAC diagram of beam No.2**

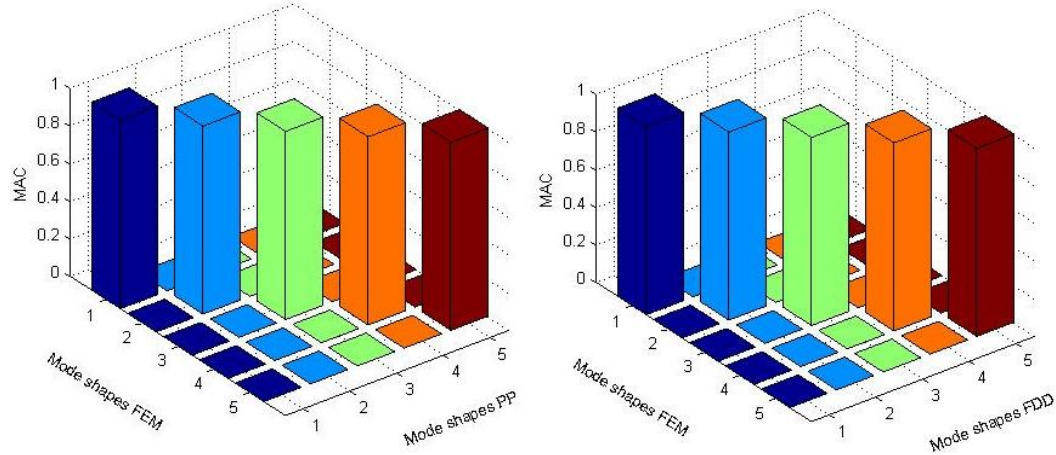

**Fig. 10- MAC diagram of beam No.3**

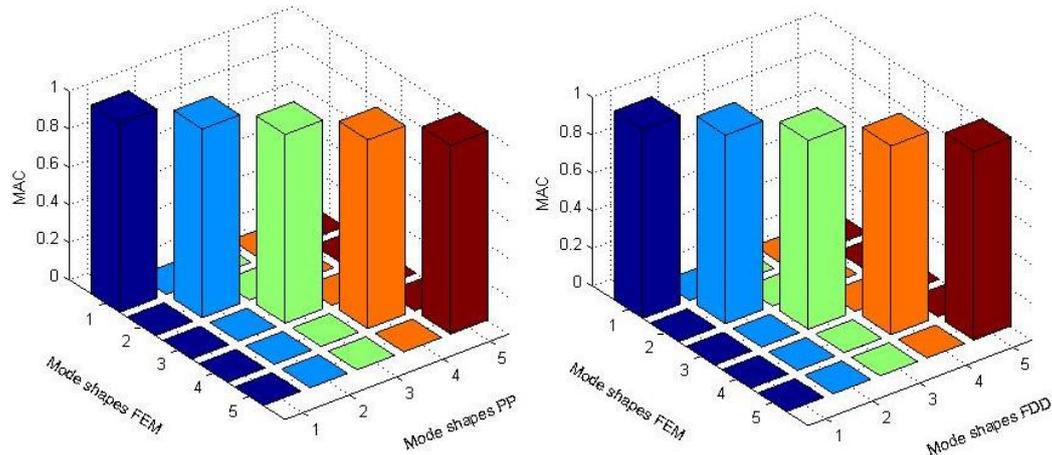

**Fig. 11- MAC diagram of beam No.4**

In the following table (tables 5-8) the values of the natural frequencies of analytical methods, FEM, PP and FDD are compared with together and the corresponding graph is drawn (Fig. 12-15).

Table 5. Natural Frequencies values for Beam No.1

| Method | Frequency 1 | Frequency 2 | Frequency 3 | Frequency 4 | Frequency 5 |
|---|---|---|---|---|---|
| **FEM** | 8.15 | 51.35 | 145.20 | 288.97 | 487.81 |
| **Analytical** | 8.15 | 51.09 | 143.09 | 280.37 | 463.49 |
| **PP** | 8.50 | 52.00 | 145.50 | 286.50 | 474.00 |
| **FDD** | 7.32 | 51.27 | 144.00 | 285.60 | 473.60 |

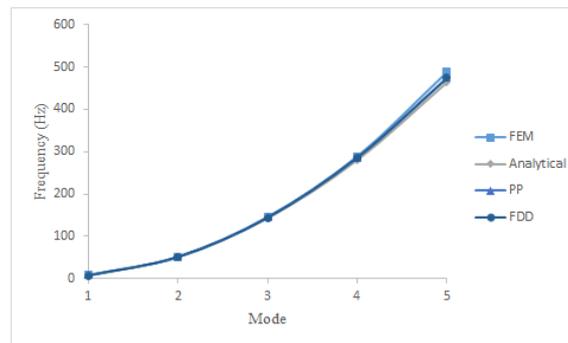

Fig. 12- Natural Frequencies for Beam No.1

Table 6. Natural Frequencies values for Beam No.2

| Method | Frequency 1 | Frequency 2 | Frequency 3 | Frequency 4 | Frequency 5 |
|---|---|---|---|---|---|
| FEM | 52.20 | 145.51 | 290.00 | 490.13 | 752.91 |
| Analytical | 51.88 | 143.02 | 280.37 | 463.50 | 692.38 |
| PP | 52.00 | 145.50 | 289.00 | 486.50 | 739.00 |
| FDD | 51.27 | 146.50 | 288.10 | 485.80 | 739.70 |

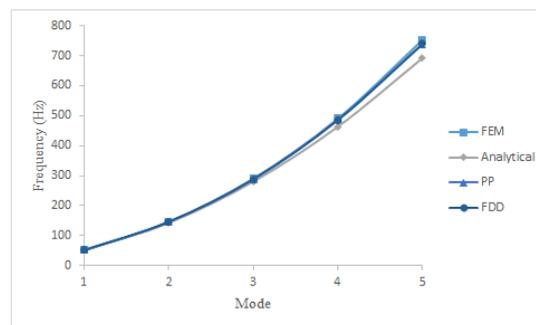

Fig. 13- Natural Frequencies for Beam No.2

Table 7. Natural Frequencies values for Beam No.3

| Method | Frequency 1 | Frequency 2 | Frequency 3 | Frequency 4 | Frequency 5 |
|---|---|---|---|---|---|
| FEM | 22.95 | 92.56 | 211.12 | 382.62 | 612.95 |
| Analytical | 22.89 | 91.55 | 205.99 | 366.21 | 572.20 |
| PP | 23.00 | 92.50 | 211.00 | 381.00 | 605.50 |
| FDD | 21.97 | 92.77 | 210.00 | 380.90 | 605.50 |

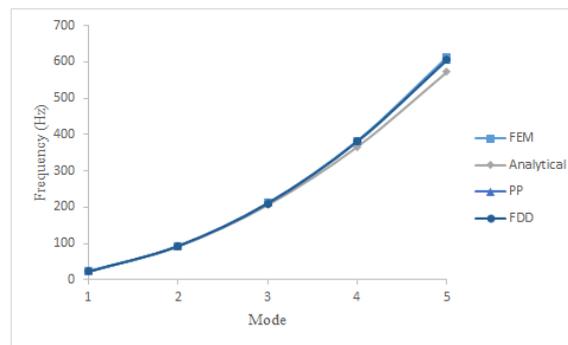

Fig. 14- Natural Frequencies for Beam No.3

Table 8. Natural Frequencies values for Beam No.4

| Method | Frequency 1 | Frequency 2 | Frequency 3 | Frequency 4 | Frequency 5 |
|---|---|---|---|---|---|
| FEM | 35.91 | 117.49 | 248.86 | 434.87 | 680.75 |
| Analytical | 35.76 | 115.89 | 241.65 | 413.42 | 630.87 |
| PP | 36.00 | 117.50 | 248.50 | 431.50 | 670.50 |
| FDD | 36.62 | 117.20 | 249.00 | 432.1 | 671.40 |

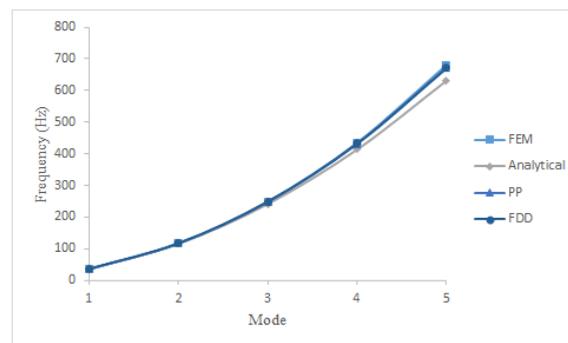

Fig. 15- Natural Frequencies for Beam No.4

## 5. Results:

This paper focused on the identification of modal properties of civil engineering structures as they vibrate in their natural environment. The main goal of the study was to evaluate and compare two of the most popular frequency-domain modal identification techniques in engineering. In this paper, modal parameters (natural frequencies, mode shapes) of the beams subjected to arbitrary loading was studied. Four beams with different boundary condition were evaluated with FDD and PP methods and then justified with the results of input-output method which was determined by frequency relation function (FRF). The comparison between the dynamic characteristics of the input-output method and output-only analysis results showed fairly accurate verification. Although very reasonable results were obtained with FDD and PP, some Natural Frequencies have consistent errors that are likely due to nonlinearities present at high levels of vibration.

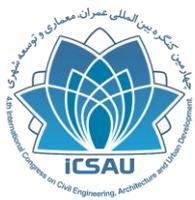
**4th. International Congress on Civil Engineering , Architecture and Urban Development**
**27-29 December 2016, Shahid Beheshti University , Tehran , Iran**